# Measuring micro-displacements of specular surfaces using speckle interferometry

**Andrés E. Dolinko[1,2,*] and Gustavo E. Galizzi[3,4]**

[1] Universidad de Buenos Aires, Facultad de Agronomía, Departamento de Ingeniería Agrícola y Uso de la Tierra, Cátedra de Física, Buenos Aires, Argentina.

[2] CONICET - Universidad de Buenos Aires, Facultad de Ciencias Exactas y Naturales, Departamento de Biodiversidad y Biología Experimental, Buenos Aires, Argentina.

[3] Instituto de Física Rosario (CONICET – UNR), Blvd. 27 de Febrero 210 bis, S2000EZP, Rosario, Argentina.

[4] Universidad Nacional de Rosario, Facultad de Ciencias Exactas, Ingeniería y Agrimensura, Av. Pellegrini 250, S2000BTP, Rosario, Argentina.

**Abstract**

The displacement field of an object surface can be measured by using speckle interferometry. This technique is based on the phenomenon of laser speckle and consists in correlating speckle interferograms taken after and before the deformation of the surface. The main requirement is that the surface under study must be optically rough to generate the speckle patterns to be correlated. In this paper, we present a very simple and intuitive method based on speckle interferometry to measure out-of-plane displacements on specular reflecting surfaces that generate no speckle patterns. The method consists in a modified digital speckle pattern interferometer which requires no special equipment other than that used in conventional speckle interferometers. The proposed method could find useful application in the measurement of thermal deformation of mirrors.

[*] Corresponding author: Andrés E. Dolinko

*E-mail address:* adolinko@df.uba.ar

## 1. Introduction

Flaw detection in mechanical components is a subject of primary importance in industry. For that reason, the development of novel nondestructive optical techniques is of great interest. In particular, the use of whole field interferometric techniques results very appropriate to test extended surfaces. In particular, The application of digital speckle pattern interferometry (DSPI) [1,2] results to be a robust technique very appropriate to evaluate micro displacement fields introduced in the sample to reveal any defect. However, in order to apply this method, the sample surface must be optically rough to be able to generate a speckle pattern.

Different interferometric approaches have been also presented to measure displacement on mirrored surfaces [3-5]. In [3], a speckle contrast measurement of the far-field diffraction pattern is presented. This approach is not exactly a displacement measurement technique. In [4], a DSPI system is introduced, although it uses the diffused field reflected by a ground glass plate, which could lead to a great absorption of light, while in [5], a holographic technique is presented, which is not based on the DSPI technique and results more sensible to spurious vibrations.

In this paper, we present an optical setup that allows measuring the displacement field of a specular surface or mirror by means of DSPI. The key feature of the proposed interferometer consists in projecting a speckle pattern previously generated by a beam expanding lens and a diffuser located in the laser light path. The setup needs no special components other than those used in DSPI interferometers, and results robust to ambience perturbations. In addition, it can be used for both specular and rough surfaces to measure out-of-plane displacements.

In the first part of the paper, we describe the optical setup used. Then, we show the sensitivity equations and displacement measurements. Finally, we present an experimental validation of the interferometer.

## 2. Description of the interferometer

The optical setup is shown in Fig. 1. It is a modification of the one described in [6,7] and consists of a conventional DSPI interferometer with out-of-plane displacement sensitivity. The key modification consists in expanding the object beam which is then passed trough a diffuser to produce a transmission speckle pattern that is projected onto the specimen.

The light of a laser source was first divided into the object and reference beams by a beam splitter ($BS_1$). The reference beam was directed to a mirror ($M_1$) linked to a piezoelectric transducer (PZT), which was controlled by an electronic unit (PCU) that was used to introduce the phase shifts needed to evaluate the phase distribution. The reference beam was then expanded by the microscope objective ($EL_1$) and was directed into the CCD camera (Pulnix TM-620) through another beam splitter ($BS_2$), where it was recombined with the speckle pattern reflected by the specimen. The video camera had a zoom lens (CL) which allows imaging a region of the specimen of approximately $20\times20mm^2$ in size. The camera output was fed into a frame grabber (Matrox Pulsar) located inside a personal computer which digitised the images in grey levels with a resolution of $512\times512$ pixels $\times$ 8 bits. The object beam is expanded by the expanding lens ($EL_3$) and then passed trough the diffuser (DFR) producing a transmission speckle pattern, that is projected onto the specular specimen. The angle of incidence and reflection of the object beams, which by the reflection law are equal ($\alpha_i=\alpha_r$), are set in such a way that the reflected light by the specular specimen (SS) is directed into the camera lens (CL), which then reach de camera CCD. The specular specimen is supported in one point by a micrometric screw (MS) that allows a controlled displacement. Fig. 2. depicts a schematics of the support specimen base designed to introduce controlled displacements.

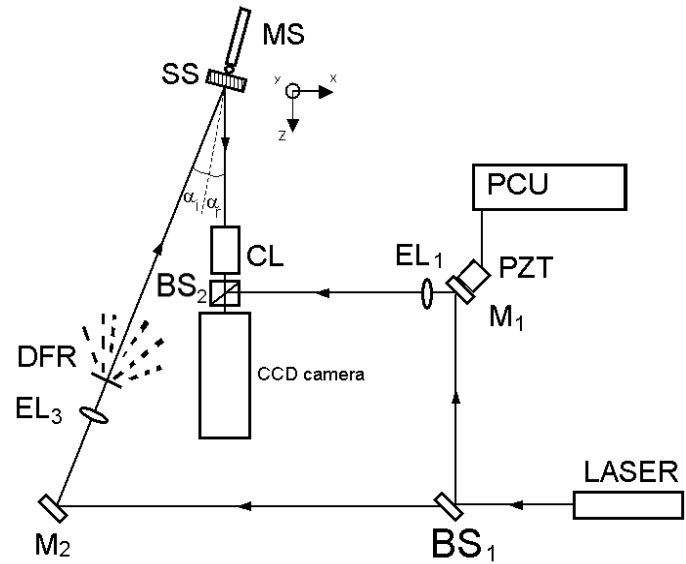

**Fig. 1**. Diagram of the experimental setup: beam splitters (BS), expanding lenses (EL), camera lens (CL), piezoelectric transducer (PZT), piezoelectric control unit (PCU), specular sample (SS), micrometric screw (MS) and diffuser (DFR).

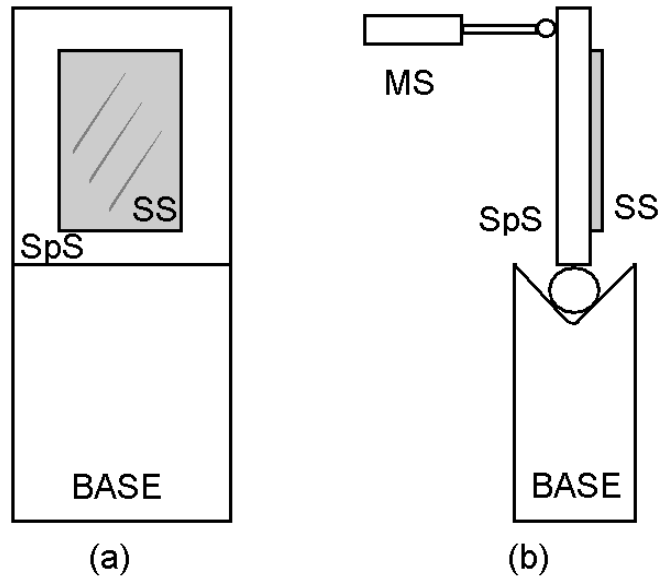

**Fig. 2**. Schematics of the specimen base: specular sample (SS), support base (SpS) and micrometric screw (MS). (a) front view and (b) lateral view.

## 3. Displacement measurement

Since we are treating with an expanded speckle field, it has different directions. Because of that, we consider the mean angle of the set of rays, that must be the angle of illumination of the laser beam before being expanded. Under this assumption, the relation between the relative phase change $\Delta\Phi(x,y)$ coded by the DSPI fringes and the out-of-plane displacement $u$ as a function of coordinates $x,y$ is given by [8]

$$\Delta\phi(x,y) = \frac{4\pi}{\lambda} u(x,y) \cos\alpha_{\mathrm{i}},$$  (1)

where $\lambda$ is the laser wavelength, and $\alpha_{\mathrm{i}}$ the mean angle of incidence and reflection of the illuminating speckle field.

Firstly, a rigid-body displacement field was introduced by tilting a specular sample and the resulting phase distribution was evaluated by the Fourier transform method [9]. Afterwards, the sample was replaced by a mirrored plate and a displacement field was introduced by a microindentation onto the specular surface. The resulting phase distribution was evaluated using the Carré phase-shifting technique [10,11] which requires two sets of four phase-shifted images before and after introducing the hole.

## 4. Experimental results and validation

In order to check the validity of the proposed technique, we evaluated the out-of-plane rigid-body displacement field that was introduced by tilting the specular sample SS around a horizontal axis with the micrometric screw MS (see Fig. 2). In this case, we used a Nd:Yag laser with wavelength $\lambda$=533nm and the angle $\alpha_{\mathrm{i}} \approx 6^{\circ}$. Then, the resulting phase distribution was computed by the Fourier transform method [9]. Fig. 3a shows the wrapped phase distribution evaluated by the Fourier transform method, obtained by tilting the sample SS towards the camera as described above. Fig. 3b depicts a 3D representation of the displacement field corresponding to the unwrapping of the phase map shown in Fig. 3a. As expected, the displacement values linearly increase as coordinate $y$ increases.

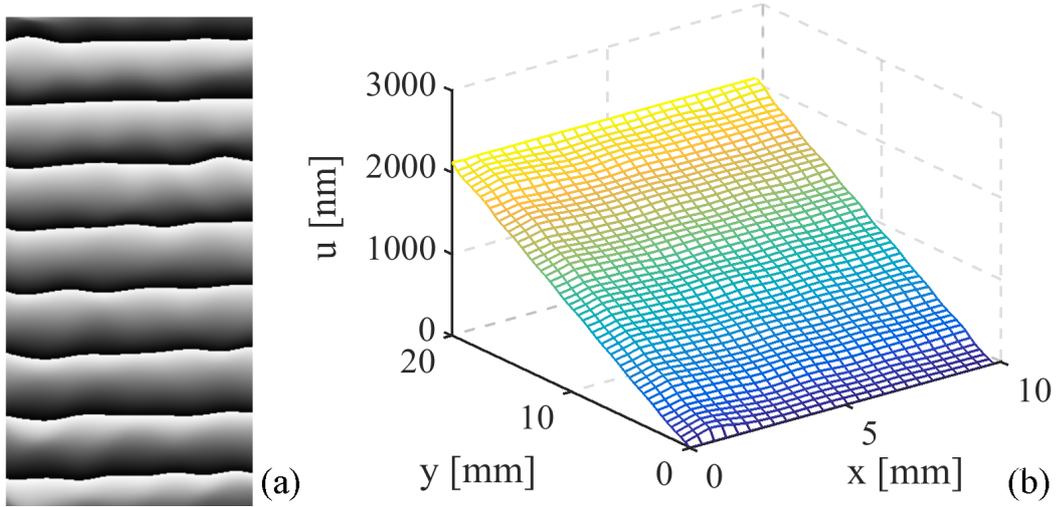

**Fig. 3**. (a) Wrapped phase distribution and (b) 3D representation of the displacement map corresponding to an out-of-plane rigid-body displacement field.

In the second test, we measured the displacement field introduced by a microindentation on a mirrored plate, following the procedure presented in [7]. The specimen to be tested was a steel plate with a rectangular cross-section and a size of $50 \times 35 \times 5$ mm$^3$. In this case, we used a He-Ne laser of $\lambda$=632 nm and the angle $\alpha_i \approx$45º. The surface of the specimen was polished to obtain a mirrored characteristic. The microindentation was introduced using a scratch tester device (Teer coatings ST30) without displacing the specimen. This device has a Rockwell C spherical diamond tip of 0.2mm of radius and a cone angle of 1201, and was located outside of the optical bench. Therefore, the specimen had to be repositioned into the same position that it had when the reference speckle interferogram was recorded. It must be noted that this procedure must be carried out with a high degree of accuracy to prevent the introduction of speckle decorrelation between both interferograms recorded before and after the introduction of the microindentation.

The repositioning was performed using a specimen holder with high stability and no moving parts [12]. The specimen rested in the holder by gravity by leaning it slightly backwards with its back surface placed against three hard metallic balls, which determined the specimen plane. Another three support pins fully determined its position. The specimen was placed with its bottom side lying on the two support pins until one of the vertical sides came to rest against the third pin.

Figure 4 shows a photograph of the specular surface of the specimen imaged by the camera where the microindentation is indicated by the white arrow. Fig. 5(a) shows the obtained wrapped phase map of the displacement field produced by the microindentation, while Fig. 5(b) depicts the corresponding unwrapped phase map. Fig. 6 shows a 3D representation of the obtained displacement field, where the orientation of $u$ axis was reversed for the sake of clarity.

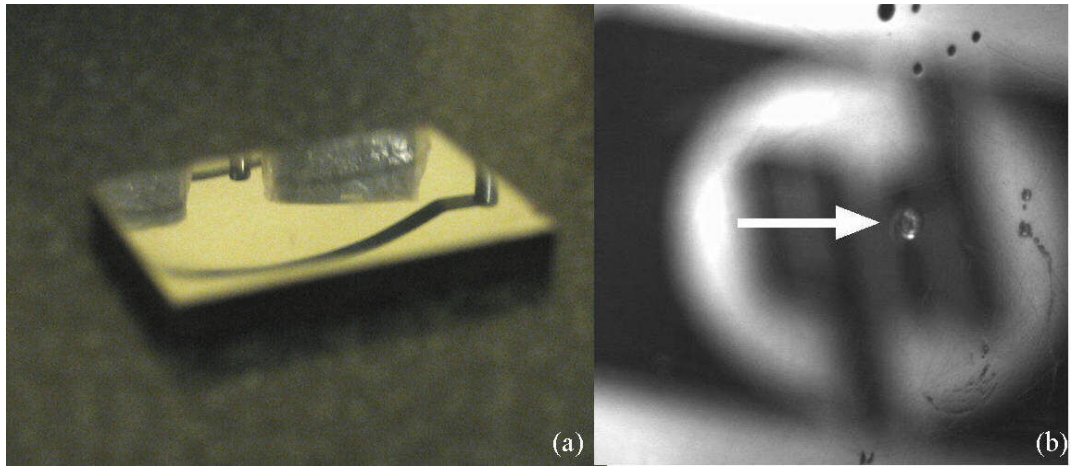

**Fig. 4**. Photograph of the (a) specimen plate and (b) the microindentation over its specular surface, indicated by the white arrow.

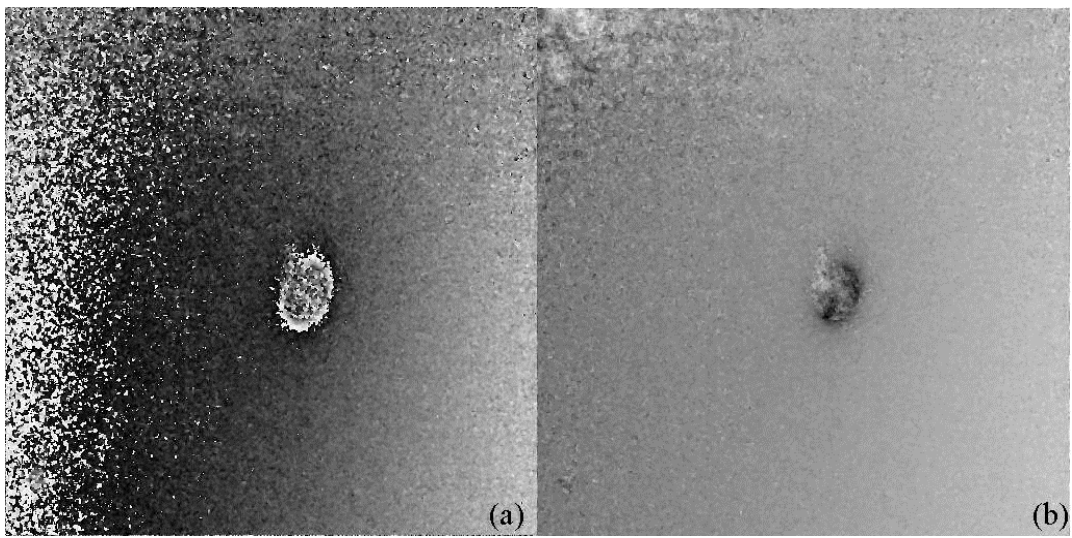

**Fig. 5**. (a) Wrapped and (b) unwrapped phase maps corresponding to the displacement field produced by the microindentation on the specular surface of the specimen.

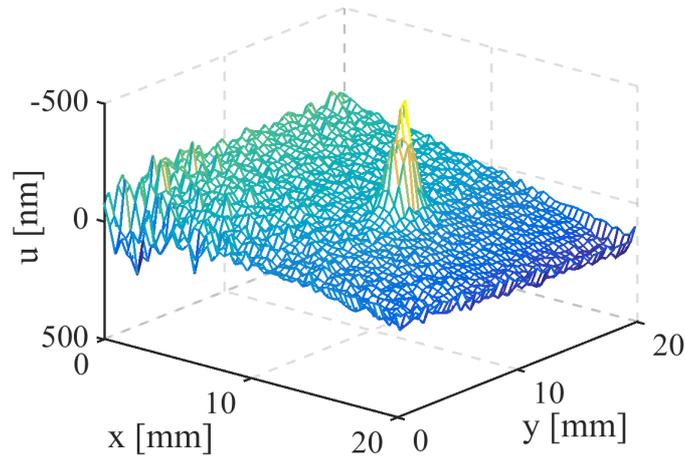

**Fig. 6**. 3D representation of the displacement field produced by the microindentation on the specular surface of the specimen.

## 5. Conclusions

This paper presents a DSPI setup able to measure displacement fields on specular surfaces (i.e. optically smooth in relation to the illumination wavelength). The system is a conventional out-of-plane speckle interferometer where a speckle pattern is introduced by means of an optic diffuser into the object beam path. That is, the required speckle information that is not generated by the specimen is previously introduced with the diffuser. A description of the setup and the equation of sensitivity are explained. Finally, we show a set of experimental results and validation of the system. The obtained results show that the proposed device is suitable to measure out-of-plane displacements in the order of the optical wavelength over specular or rough surfaces.

*Acknowledgements*

The authors would like to thank the financial support provided by the Consejo Nacional de Investigaciones Científicas y Técnicas [PIP11220150100607]; and Universidad de Buenos Aires [UBACyT2002015020019].